\documentclass[prd,aps,twocolumn,showpacs,preprintnumbers,amsmath,amssymb,nofootinbib]{revtex4}
\usepackage[dvips]{graphicx}
\usepackage{epsf}
\usepackage{amsmath}
\usepackage{amssymb}
\usepackage{graphicx}% Include figure files
\usepackage{dcolumn}% Align table columns on decimal point
\usepackage{bm}% bold math
\begin{document}

\title{Testing the interaction between dark energy and dark matter via latest observations}

\author{Jian-Hua He, Bin Wang}   \affiliation{INPAC and Department of Physics, Shanghai Jiao Tong University, 200240 Shanghai, China}
\author{Elcio Abdalla}
\affiliation{Instituto de Fisica, Universidade de Sao Paulo, CP 66318, 05315-970, Sao Paulo, Brazil}

\begin{abstract}
Cosmological analysis based on currently available observations are unable to rule out a sizeable coupling between dark energy and dark matter.
However, the signature of the coupling is not easy to grasp, since the coupling is degenerate with other cosmological parameters, such as the
dark energy equation of state and the dark matter abundance. We discuss the possible ways to break such degeneracy. Based on the perturbation
formalism, we carry out the global fitting by using latest observational data and get a tight constraint on the interaction between dark sectors.
We find that the appropriate interaction can alleviate the coincidence problem.
\end{abstract}

 \maketitle
\section{Introduction}
There has been convincing evidence indicating that our universe is composed of nearly $25\%$ cold dark matter (DM) plus a small fraction of
baryonic matter and around $70\%$ dark energy (DE)\cite{1}. One leading candidate of such a DE is the cosmological constant, representing a
vacuum energy density with constant equation of state $w=-1$. However, it is difficult to understand such a cosmological constant in terms of
fundamental physics. Its observed value is far below that estimated in quantum field theory, what is referred to as the cosmological constant
problem. Moreover, using the cosmological constant to explain the DE, there is no natural understanding why the constant vacuum energy and matter
energy densities are precisely of the same order today. This is the so-called coincidence problem.

Considering that DE and DM are dominant sources of the content of the universe, it is natural, in the framework of field theory, to consider the
inevitable interaction between them \cite{secondref}. An appropriate interaction between DE and DM can provide a mechanism to alleviate the
coincidence problem \cite{10}-\cite{14}. A non-minimal coupling in dark sectors can affect significantly the expansion history of the universe
and the density perturbation evolution, changing the growth history of cosmological structures. The possibility that DE and DM interact with each
other has been widely discussed recently \cite{10}-\cite{AbdallaPLB09}.  A number of studies have been devoted to analyze the constraints on the
dark sectors mutual interaction from the probes of the cosmic expansion history by using the WMAP, SNIa, BAO and SDSS data etc
\cite{71}-\cite{pp}. Interestingly it was disclosed that the late ISW effect has the unique ability to provide insight into the coupling between
dark sectors \cite{hePRD09}.

Furthermore, complementary probes of the coupling within the dark sectors have been carried out in the study of the growth of cosmic structure
\cite{31}-\cite{AbdallaPLB09}. It was found that a non-zero interaction between dark sectors leaves a clear change in the growth index
\cite{31,Caldera}. In addition, it was suggested that the dynamical equilibrium of collapsed structures such as clusters would acquire a
modification due to the coupling between DE and DM \cite{pt,AbdallaPLB09}. Comparing the naive virial masses of a large sample of clusters with
their masses estimated by X-ray and by weak lensing data, a small positive coupling has been tightly constrained \cite{AbdallaPLB09}, which
agrees with the results given in \cite{hePRD09} from CMB. The small positive coupling indicates that there is energy transfer from DE to DM,
which can help to alleviate the coincidence problem \cite{hePRD09,heJCAP08}.

Both DE and DM are currently only detected via their gravitational effects and any change in the DE density is conventionally attributed to its
equation of state $w$. This  status leads to an inevitable degeneracy while extracting the signature of the  interaction between dark sectors and
other cosmological parameters. In this work, we will first discuss the degeneracy between the DE and DM coupling and the equation of state (EoS)
of DE in the background dynamics. Furthermore, we will extend our discussion to the perturbed spacetime by considering the perturbation evolution
of DE and DM. We review the formalism of the perturbation theory when there is an interaction between dark sectors. Based upon this formalism we
explore the possibility of breaking the degeneracy between the coupling and other cosmological parameters, such as the EoS parameter $w$ of DE as
well as the DM abundance. This can help us extract a tighter constraint on the interaction between dark sectors from observations.

\section{The back ground dynamics \label{background}}
In the spatially flat Friedmann-Robertson-Walker(FRW) background, if there is an interaction between DE and DM, neither of them can evolve
independently. The (non)conservation equations are described by
\begin{eqnarray}
&&\rho'_c+3\mathcal{H}\rho_c = aQ_c\nonumber \\
&&\rho'_d+3\mathcal{H}(1+w)\rho_d = aQ_d\quad ,
\end{eqnarray}
where the subscript ``c" represents DM and ``d" stands for DE. $Q_{\lambda}$ is the term leading to energy transfer. Considering that there is
only energy transfer between DE and DM, we have $Q_c=-Q_d=Q$. The sign of $Q$ determines the direction of energy transfer. For positive $Q$, the
energy flows from DE to DM. For negative $Q$, the energy flow is reversed. Since we know neither the physics of DM nor that of DE at the present
moment, we cannot write out the precise form of the interaction between them from first principles (see, however, \cite{secondref}). One has to
specify the interaction either from the outset \cite{11}, or determine it from phenomenological requirements \cite{Olivares, heJCAP08}.

For the sake of generality, we consider the phenomenological description of the interaction between DE and DM as a linear combination of energy
densities of dark sectors $Q=3H(\xi_1\rho_m+\xi_2\rho_d)$ \cite{hePLB09, 31}. In studying the curvature perturbation it has been made clear that
when the interaction is proportional to the energy density of DE ($Q=3H\xi_2\rho_d$), we get a stable curvature perturbation; however, when the
interaction is proportional to the DM density ($Q=3H\xi_1\rho_m$) or total dark sectors ($Q=3H\xi(\rho_m+\rho_d)$), the curvature perturbation
can only be stable when the constant DE EoS satisfies $w<-1$ \cite{hePLB09}. For the case of a time-dependent DE EoS, the stability of curvature
perturbations was discussed in \cite{stability}.

The presence of the coupling also changes the DM and DE redshift dependence acting as an extra contribution to their effective equation of state.
Indeed, the effective background equations of state for the two fluids are
\begin{eqnarray}
w_{{\rm c,eff}}&=&-\frac{aQ_c}{3\mathcal{H}\rho_c}\nonumber \\
w_{{\rm d,eff}}&=&w-\frac{aQ_d}{3\mathcal{H}\rho_d}\quad .
\end{eqnarray}
Choosing different forms for the interaction, the effective background EoS can be read from table I. For convenience, we label our models with
Roman numbers.
\begin{table*}
\caption{Coupling models}\label{EoS}
\begin{tabular}{|c|c|c|c|c|c|}
\hline
\hline
Model&$Q$ & DE EoS&$w_{{\rm c,eff}}$ & $w_{{\rm d,eff}}$ & Constrains \\
\hline
I & $3\xi_2H\rho_d(\xi_1=0)$& $-1<w<0$&$-\xi_2/r$ & w+$\xi_2$ & $\xi_2<-2w\Omega_c$  \\
\hline
II & $3\xi_2H\rho_d(\xi_1=0)$&$w<-1$& $-\xi_2/r$ & w+$\xi_2$ & $\xi_2<-2w\Omega_c$\\
\hline
III & $3\xi_1H\rho_c(\xi_2=0)$&$w<-1$& $-\xi_1$ & $w+\xi_1r$ &$0<\xi_1<-w/4$ \\
\hline
IV & $ 3\xi H(\rho_c + \rho_d)(\xi=\xi_1=\xi_2)$ &$w<-1$& $-\xi(1+1/r)$ & $ w+\xi(r+1)$& $0<\xi<-w/4$ \\
\hline
\hline
\end{tabular}\label{table}
\end{table*}

We define $r=\rho_{c}/\rho_d$ as the ratio of the energy densities of DM and DE. In order to solve the coincidence problem we require the ratio
of $r$ to be a constant in the expansion history of our universe. This leads to the condition
$r'=\frac{\rho_{c}'}{\rho_d}-r\frac{\rho_d'}{\rho_d}=0$, which yields a quadratic equation,
\begin{equation}
\xi_1r^2+(\xi_1+\xi_2+w)r+\xi_2=0\label{quadratic}\quad .
\end{equation}
When the coupling is proportional to the dark energy density (Models I, II)  $\xi_1=0$, Eq.~(\ref{quadratic}) has only one root,
$r=-\frac{\xi_2}{\xi_2+w}$, which will appear in the future. In \cite{heJCAP08} it was found that when the coupling is proportional to the energy
density of DE, $\xi_2>-2w\Omega_c$ leads to a negative energy density of cold DM($\rho_{\rm{c}}<0$) in the past, which is unphysical. We thus
require $\xi_2<-2w\Omega_c$. In such a case, the effective DE EoS  reads $w_{\rm {d,eff}}=w+\xi_2$, which shows that the coupling and the DE EoS
are entangled and it is impossible to distinguish the coupling from DE EoS by investigating the background evolution.

When the coupling is proportional to the dark matter density or to the total dark sector (Models III, IV),  there are two roots of the quadratic
equation Eq.~(\ref{quadratic}),
\begin{eqnarray}
(r\xi_1)_1&=&-\frac{1}{2}(w+\xi_1+\xi_2)+\frac{1}{2}\sqrt{(\xi_1+\xi_2+w)^2-4\xi_1\xi_2}\nonumber \\
(r\xi_1)_2&=&-\frac{1}{2}(w+\xi_1+\xi_2)-\frac{1}{2}\sqrt{(\xi_1+\xi_2+w)^2-4\xi_1\xi_2}\quad \nonumber.
\end{eqnarray}
The first root happens in the past and the second one happens  in the future.  For the interaction proportional to the dark matter density (Model
III) ($\xi_2=0$), when $\xi_1<-w/4$, the roots are real and can be simplified. We have,
\begin{eqnarray}
(r\xi_1)_1&=&-(w+\xi_1)>0\nonumber \\
(r\xi_1)_2&=&0
\end{eqnarray}
The effective DE EoS can behave differently in the past and in  the future. In  the early time of the universe, the effective DE EoS can be of
the form
\begin{equation}
w_{{\rm d,eff}}\approx-\xi_1\quad .
\end{equation}
We learn that the coupling is entangled with the DE EoS. When the  universe evolves to the present time or into the future, the effective DE EoS
reads
\begin{equation}
w_{{\rm d,eff}}\approx w
\end{equation}
which does not depend on the coupling and thus the degeneracy between  DE EoS and coupling no longer exists in the background dynamics.

When the interaction is proportional to the energy density of total dark  sectors (Model IV), ($\xi_1=\xi_2$), Eq.~(\ref{quadratic}) has two real
roots when $\xi<-w/4$,
\begin{eqnarray}
(r\xi)_1&=&-\frac{1}{2}(w+2\xi)+\frac{1}{2}\sqrt{(2\xi+w)^2-4\xi^2}\nonumber \\
        &\approx&-(w+2\xi)\nonumber \\
(r\xi)_2&=&-\frac{1}{2}(w+2\xi)-\frac{1}{2}\sqrt{(2\xi+w)^2-4\xi^2}\nonumber \\
        &\approx&-\frac{\xi^2}{w}\approx\xi^2,
\end{eqnarray}
according to which the effective DE EoS turns out to be $w_{{\rm d,eff}}\approx-\xi$ in the early time of the universe and $w_{{\rm
d,eff}}\approx w+\xi^2+\xi$ at late times. There is a degeneracy between the coupling between dark sectors and the DE EoS in the background
dynamics. From the background dynamics we see that when we introduce the interaction  between DE and DM, it is possible to have the scaling
solution of the ratio between DM and DE, which can help to alleviate the coincidence problem. However, in the background dynamics there appears
an inevitable degeneracy between the coupling between dark sectors and the DE EoS. In general this degeneracy cannot be broken by just
investigating the dynamics of the background spacetime, except in the case when the coupling is proportional to the dark matter density (Model
III). In the following we are going to explore the possibility of breaking the degeneracy between the coupling and other cosmological parameters
in the perturbed spacetime by considering the perturbation evolution of DE and DM.

\section{The perturbation formalism \label{perturbation}}
In  this section, we will go over the first order metric perturbation theory in the presence of coupling between DE and DM. The perturbed
space-time at first order reads
\begin{eqnarray}
ds^2 &=& a^2[-(1+2\psi)d\tau^2+2\partial_iBd\tau
dx^i\nonumber \\
&+&(1+2\phi)\delta_{ij}dx^idx^j+D_{ij}Edx^idx^j],\label{perturbedspacetime}
\end{eqnarray}
where $\psi, B, \phi, E$ represent the scalar metric perturbations,
$a$ is the cosmic scale factor and
\begin{equation}
D_{ij}=(\partial_i\partial_j-\frac{1}{3}\delta_{ij}\nabla^2).
\end{equation}
We work with the energy-momentum tensor
\begin{equation}
T^{\mu\nu}=\rho U^{\mu}U^{\nu}+p(g^{\mu\nu}+U^{\mu}U^{\nu}),
\end{equation}
for  a two-component system consisting of DE and DM. The covariant description of the energy-momentum transfer between DE and DM is given by
\cite{Sasaki}
\begin{eqnarray}
\nabla_{\mu}{T_{(\lambda)}}^{\mu\nu}={Q_{(\lambda)}}^{\nu}\label{energy-stress}
\end{eqnarray}
where ${Q_{(\lambda)}}^{\nu}$ is a four vector governing the  energy-momentum transfer between different components. The subindex $\lambda$
refers to DM and DE respectively. For the whole system, the energy and momentum are conserved, and the transfer vector satisfies
$$\sum_{\lambda}{Q_{(\lambda)}}^{\nu}=0\quad .$$
In the Fourier space the covariant form of perturbed Eq.~(\ref{energy-stress}) reads \cite{hePLB09,31}
\begin{widetext}
\begin{eqnarray}
\delta_{\lambda}'+3\mathcal{H}(\frac{\delta p_{\lambda}}{\delta
\rho_{\lambda}}-w_{\lambda})\delta_{\lambda}+(1+w_{\lambda})kv_{\lambda}&=&
-3(1+w_{\lambda})\phi'+(2\psi-\delta_{\lambda})\frac{a^2Q^0_{\lambda}}{\rho_{\lambda}}+\frac{a^2
\delta Q^0_{\lambda}}{\rho_{\lambda}},\\\label{perturbed}
(v_{\lambda}+B)'+\mathcal{H}(1-3w_{\lambda})(v_{\lambda}+B) -\frac{k}{1+w_{\lambda}}\frac{\delta p_{\lambda}}{\delta
\rho_{\lambda}}\delta_{\lambda}&=&
-\frac{w_{\lambda}'}{1+w_{\lambda}}(v_{\lambda}+B)+k\psi
-\frac{a^2Q^0_{\lambda}}{\rho_{\lambda}}v_{\lambda}-\frac{w_{\lambda}a^2Q^0_{\lambda}}{(1+w_{\lambda})\rho_{\lambda}}B+\frac{a^2\delta
Q_{p\lambda}}{(1+w_{\lambda})\rho_{\lambda}}\nonumber.\\
\end{eqnarray}
\end{widetext}
By constructing gauge invariant quantities \cite{hePRD09},
\begin{widetext}
\begin{eqnarray}
\Psi&=& \psi -
\frac{1}{k}\mathcal{H}(B+\frac{E'}{2k})-\frac{1}{k}(B'+\frac{E^{''}}{2k}),\quad \Phi=\phi+\frac{1}{6}E-\frac{1}{k}\mathcal{H}(B+\frac{E'}{2k}),\quad \delta Q_{p\lambda}^{I}=\delta Q_{p\lambda }-Q^0_{\lambda
}\frac{E'}{2k};\nonumber\\
Q^{0I}_{\lambda }&=&\delta
Q^0_{\lambda}-\frac{Q^{0'}_{\lambda}}{\mathcal{H}}(\phi+\frac{E}{6})+Q^{0}_{\lambda}\left[\frac{1}{\mathcal{H}}(\phi+\frac{E}{6})\right]',\quad
D_{g\lambda}=
\delta_{\lambda}-\frac{\rho_{\lambda}'}{\rho_{\lambda}\mathcal{H}}\left(\phi+\frac{E}{6}\right),\quad
V_{\lambda} = v_{\lambda} -\frac{E'}{2k},
\end{eqnarray}
\end{widetext}
we obtain the general gauge-invariant perturbation equations for DM and DE
respectively,
\begin{widetext}
\begin{eqnarray}
D'_{gc}+\left \{ \left(\frac{a^2Q_c^0}{\rho_c\mathcal{H}}\right)'+\frac{\rho_c'}{\rho_c\mathcal{H}}\frac{a^2Q_c^0}{\rho_c} \right \}\Phi +
\frac{a^2 Q_c^0}{\rho_c}D_{gc}+ \frac{a^2Q_c^0}{\rho_c\mathcal{H}}\Phi' &=& -kU_c +2\Psi \frac{a^2Q_c^0}{\rho_c}+\frac{a^2\delta
Q_c^{0I}}{\rho_c}+\frac{a^2Q_c^{0'}}{\rho_c\mathcal{H}}\Phi-\frac{a^2Q_c^0}{\rho_c}\left(\frac{\Phi}{\mathcal{H}}\right)',\nonumber
\\
U_c'+\mathcal{H}U_c&=&k\Psi-\frac{a^2Q_c^0}{\rho_c}U_c+\frac{a^2\delta Q_{pc}^{I}}{\rho_c}\label{Gdc};
\end{eqnarray}
\begin{eqnarray}
&&D'_{gd}+\left\{\left(\frac{a^2Q_d^0}{\rho_d\mathcal{H}}\right)'-3w'+3(C_e^2-w)\frac{\rho_d'}{\rho_d}+\frac{\rho_d'}{\rho_d\mathcal{H}}\frac{a^2Q_d^0}{\rho_d}\right\}\Phi+\left\{3\mathcal{H}(C_e^2-w)+\frac{a^2Q_d^0}{\rho_d}\right\}D_{gd}+\frac{a^2Q_d^0}{\rho_d\mathcal{H}}\Phi'\nonumber\\
&=&-kU_d+3\mathcal{H}(C_e^2-C_a^2)\frac{\rho_d'}{\rho_d}\frac{U_d}{(1+w)k}+2\Psi\frac{a^2Q_d^0}{\rho_d}+\frac{a^2\delta
Q_d^{0I}}{\rho_d}+\frac{a^2Q_d^{0'}}{\rho_d\mathcal{H}}\Phi-\frac{a^2Q_d^0}{\rho_d}\left(\frac{\Phi}{\mathcal{H}}\right)'\nonumber \label{DEU},\\
&&U_d'+\mathcal{H}(1-3w)U_d=kC_e^2D_{gd}+kC_e^2\frac{\rho_d'}{\rho_d\mathcal{H}}\Phi-\left(C_e^2-C_a^2\right)\frac{U_d}{1+w}\frac{\rho_d'}{\rho_d}+(1+w)k\Psi
-\frac{a^2Q_d^0}{\rho_d}U_d+(1+w)\frac{a^2\delta
Q_{pd}^{I}}{\rho_d}\label{Gdd}.\nonumber\\
\end{eqnarray}
\end{widetext}
where $D_{g\lambda}$ is the energy density contrast in spatial flat  gauge\cite{Sasaki}, $U_{\lambda}=(1+w_{\lambda})V_{\lambda}$ and
$V_{\lambda}$ is the gauge invariant peculiar velocity. We have employed
\begin{equation}
\frac{\delta p_d}{\rho_d} = C_e^2\delta_d-(C_e^2-C_a^2)\frac{\rho_d'}{\rho_d}\frac{v_d+B}{k},
\end{equation}
where $C_e^2$ is the effective sound speed  of DE at its rest frame which is gauge invariant under gauge transformation and $C_a^2$ is the
adiabatic sound speed.

Eqs.(\ref{Gdc}) and (\ref{Gdd}) are the most generic form regardless of the detailed description of the interaction ${Q_{(\lambda)}}^{\nu}$.

The four vector ${Q_{(\lambda)}}^{\nu}$ can be phenomenologically decomposed into two parts with respect to a given observer $\lambda'$ with four
velocity ${U_{(\lambda')}}^{\mu}$.
\begin{equation}
{Q_{(\lambda)}}^{\mu}=Q{_{(\lambda \lambda')}}{U_{(\lambda')}}^{\mu}+{F_{(\lambda \lambda')}}^{\mu}\quad ,
\end{equation}
where $Q{_{(\lambda \lambda')}}=-{U_{(\lambda')}}_{\nu}Q{_{(\lambda)}}^{\nu}$  is the energy transfer rate of $\lambda$ component observed by
$\lambda'$ observer. ${F_{(\lambda \lambda')}}^{\mu}={{h_{(\lambda')}}^{\mu}}_{\nu}{Q_{(\lambda)}}^{\nu}$ is the corresponding momentum transfer.
In appendix~\ref{appendix} we show that such decomposition of ${Q_{(\lambda)}}^{\nu}$ and its perturbed form are identities regardless of the
observer. The four vector ${Q_{(\lambda)}}^{\nu}$ must be specified directly, according to the physical meaning. As discussed in~\ref{mes}, in
cosmology, we need to specify the coupling vector  ${Q_{(\lambda)}}^{\nu}$ in the co-moving frame as
\begin{eqnarray}
{Q_{(\lambda)}}^{\nu}&=&\left[\frac{Q_{(\lambda)}}{a},0,0,0\right]^{T}\label{couplev}
\end{eqnarray}
$Q_{(\lambda)}$ is the module of four vector ${Q_{(\lambda)}}^{\nu}$. The perturbed form $\delta Q_{(\lambda)}^0$  can be uniquely determined
from the background energy-momentum transfer $Q_{(\lambda)}^{\mu}$. From
\begin{equation}
{Q_{(\lambda)}}^{\mu}Q_{(\lambda)\mu}=g_{00}({Q_{(\lambda)}}^0)^2=-{Q_{(\lambda)}}^2
\end{equation}
where $Q_{(\lambda)}=a{Q_{(\lambda)}}^{0}$ is a scalar in the FRW space  and the minus sign indicates that ${Q_{(\lambda)}}^{\mu}$ is time-like,
we can obtain the perturbation form
\begin{equation}
\delta Q_{(\lambda)}^0=-\frac{\psi}{a} Q_{(\lambda)} + \frac{1}{a}\delta Q_{(\lambda)}\quad .
\end{equation}
It can be shown that the zero component of the perturbed energy-momentum transfer is covariant. The spatial component of the perturbed
energy-momentum transfer $\delta Q^i_{(\lambda)}$ is independent of the zeroth component.  It refers to the non-gravitational force and is
composed of two parts,
\begin{eqnarray}
\delta Q_{p\lambda}=\delta Q_{p\lambda}^I\mid_{t}+Q_{(\lambda)}^0v_{t}\label{deltaQ}
\end{eqnarray}
where $\delta Q_{p\lambda}$ is the potential of the perturbed energy-momentum transfer $\delta Q^i_{(\lambda)}$, $\delta Q_{p\lambda}^I\mid_{t}$
is the external non-gravitational force density and $v_{t}$ is the average velocity of the energy transfer. $v_{t}$ is a free quantity which
needs to be specified according to physics. In \cite{maartens}, $v_{t}$ was allowed to follow the peculiar velocity of DM or DE respectively. If
we allow $v_{t}=v_c$ or $v_{t}=v_d$, we can reproduce the result in \cite{maartens}. In our analysis, we consider the fact that there is no
non-gravitational interaction in the DE and DM coupled system, only the inertial drag effect appears in the system due to the stationary energy
transfer between DE and DM as discussed in \cite{peacock}. Thus we set $v_{t}=0$ and $\delta Q_{p\lambda}^I\mid_{t}=0$, which leads to the vanish
of $\delta Q^i$.

In constructing the four vector Eq.~(\ref{couplev}), the module $Q_{\lambda}$ can be chosen as any combinations of scalar in the FRW space, such
as the energy density ${\rho_{(\lambda)}}={T_{(\lambda)}}^{\mu \nu}U_{(\lambda)\mu}U_{(\lambda)}\nu$, expansion
$H_{(\lambda)}=\frac{1}{3}\nabla_{\mu}{U_{(\lambda)}}^{\mu}$, or any scalar function thereof. Considering that $Q_{\lambda}$ is observer
independent and so does the energy density and its perturbed form, we require $H$ to be a global quantity to avoid the ambiguity of the observer
dependence. In general phenomenological description, we can assume the coupling to be given by
\begin{eqnarray}
Q_{c}&=&Q=3H(\xi_1\rho_c+\xi_2\rho_d)\nonumber \\
Q_{d}&=&-Q=-3H(\xi_1\rho_c+\xi_2\rho_d)\quad .\nonumber
\end{eqnarray}
The perturbed forms read
\begin{eqnarray}
\delta Q_{c}&=&3H(\xi_1\delta\rho_c+\xi_2\delta\rho_d)\nonumber \\
\delta Q_{d}&=&-3H(\xi_1\delta\rho_c+\xi_2\delta\rho_d)\nonumber \\
\delta Q_{c}^0 &=& -3H(\xi_1\rho_c+\xi_2\rho_d)\frac{\psi}{a}+3H(\xi_1\delta \rho_c + \xi_2 \delta \rho_d)\frac{1}{a}\nonumber\\
\delta Q_{d}^0 &=& 3H(\xi_1\rho_c+\xi_2\rho_d)\frac{\psi}{a}-3H(\xi_1\delta \rho_c + \xi_2 \delta \rho_d)\frac{1}{a}.\nonumber
\end{eqnarray}
The first terms in the last two equations were omitted in our previous work \cite{hePRD09}, however, only slightly modification will be brought
by adding these terms.

The gauge invariant quantities $\delta Q_c^{0I}$ and $\delta Q_d^{0I}$ appeared in Eqs.~(\ref{Gdc}),~(\ref{Gdd}) as defined in \cite{hePRD09} can
be calculated as,
\begin{widetext}
\begin{eqnarray}
\frac{a^2\delta Q_c^{0I}}{\rho_c}&=&-3\mathcal{H}(\xi_1+\xi_2/r)\Psi+3\mathcal{H}\{\xi_1D_{gc}
+\xi_2D_{gd}/r\}\Phi+3(\xi_1\frac{\rho_c'}{\rho_c}+\frac{\xi_2}{r}\frac{\rho_d'}{\rho_d})
-\frac{a^2}{\rho_c}\frac{Q_c^{0'}}{\mathcal{H}}\Phi+\frac{a^2Q_c^0}{\rho_c}\left[
\frac{\Phi}{\mathcal{H}}\right]'\nonumber \\
\frac{a^2\delta Q_d^{0I}}{\rho_d}&=&3\mathcal{H}(\xi_1r+\xi_2)\Psi-3\mathcal{H}\{\xi_1D_{gc}r+\xi_2D_{gd}\}
-3(\xi_1r\frac{\rho_c'}{\rho_c}+\frac{\rho_d'}{\rho_d}\xi_2)\Phi-\frac{a^2Q_d^{0'}}{\rho_d}\Phi+\frac{a^2Q_d^0}{\rho_d}\left[
\frac{\Phi}{\mathcal{H}}\right]'\nonumber
\end{eqnarray}
\end{widetext}
where $r=\rho_c/\rho_d$. Inserting the above expressions into Eqs.~(\ref{Gdc}),~(\ref{Gdd}) and neglecting the spatial perturbations  $\delta
Q_{p\lambda}^{I}=0$, we obtain the general gauge-invariant perturbation equations for DM and DE respectively.
\begin{widetext}
\begin{eqnarray}
D_{gc}'&=&-kU_c+3\mathcal{H}\Psi(\xi_1+\xi_2/r)-3(\xi_1+
\xi_2/r)\Phi'+3\mathcal{H}\xi_2(D_{gd}-D_{gc})/r \nonumber\quad ,\\
U_c'&=&-\mathcal{H}U_c+k\Psi-3\mathcal{H}(\xi_1+\xi_2/r)U_c \label{cvelocity}\quad ,\\
D_{gd}' & = & -3\mathcal{H}(C_e^2-w)D_{gd}+\left\{3w'-9\mathcal{H}(w-C_e^2)\left(\xi_1r+\xi_2+
1+w\right )\right\}\Phi
\nonumber\\
&-&9\mathcal{H}^2(C_e^2-C_a^2)\frac{U_d}{k}+3(\xi_1r+\xi_2)\Phi'-3\Psi\mathcal{H}(\xi_1
r+\xi_2)+3\mathcal{H}\xi_1r(D_{gd}-D_{gc})\nonumber \\
&-&9\mathcal{H}^2(C_e^2-C_a^2)(\xi_1r+\xi_2)\frac{U_d}{(1+w)k}-kU_d \nonumber\\
U_d'& = &-\mathcal{H}(1-3w)U_d-3kC_e^2\left(\xi_1
r+\xi_2+1+w\right )\Phi+3\mathcal{H}(C_e^2-C_a^2)(\xi_1r+\xi_2)\frac{U_d}{(1+w)} \nonumber\\
&+&3(C_e^2-C_a^2)\mathcal{H}U_d+kC_e^2D_{gd}+(1+w)k\Psi+3\mathcal{H}
(\xi_1r+\xi_2)U_d. \label{dvelocity}
\end{eqnarray}
\end{widetext}
The general gauge invariant formalism fully removes the ambiguity of gauge choice.  However, it can be solved through gauge-dependent methods by
picking a peculiar gauge, without loosing generality (see.\cite{Sasaki} chapter III for details). The results will be the same for different
gauges if the gauge is fully fixed (see appendix ~\ref{gauge}). In the following discussion we will choose  the Conformal Newtonian gauge
together with the adiabatic initial conditions as we used in \cite{hePLB09}.

\begin{figure*}
\centering
\includegraphics[width=2.8in,height=2in]{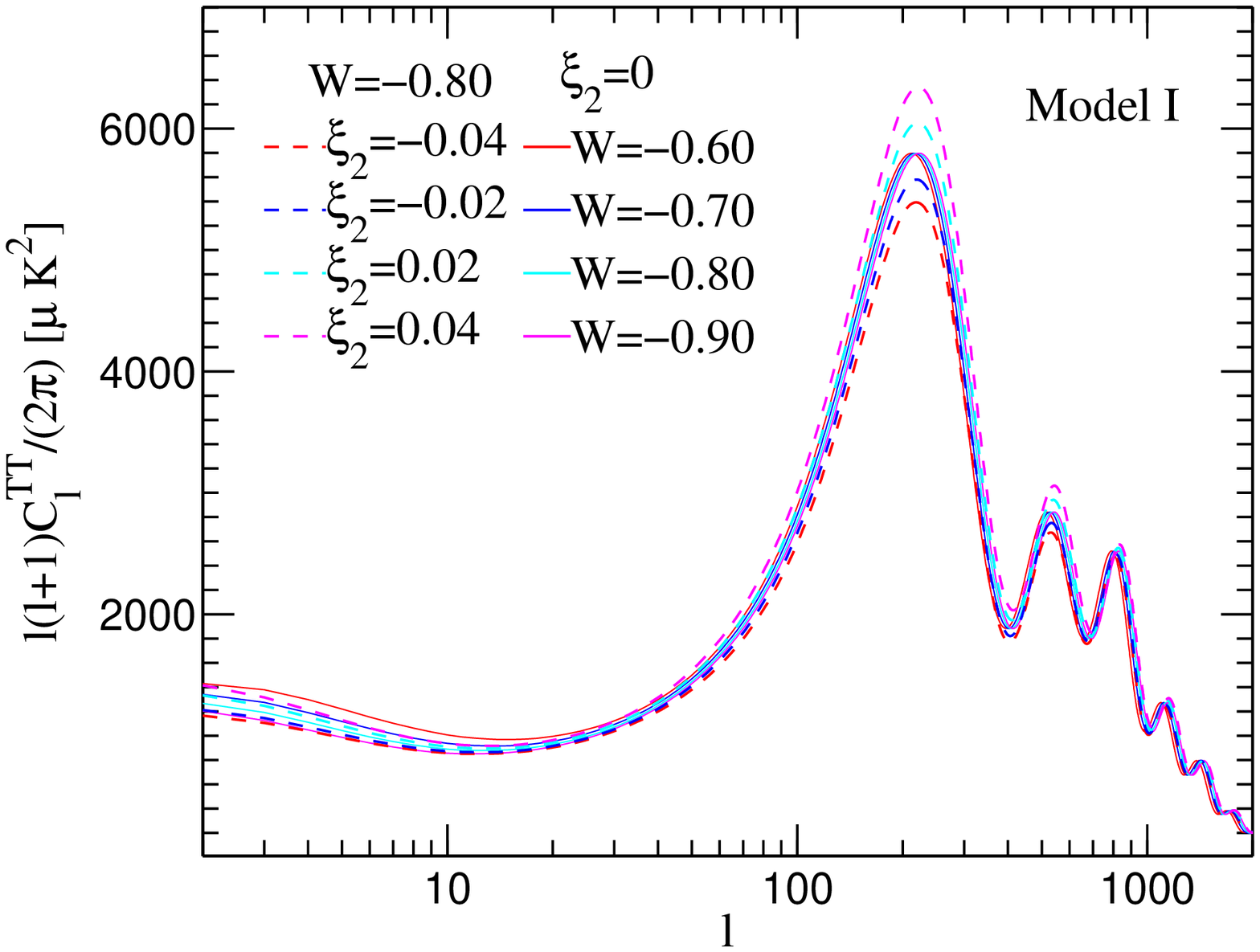}
\includegraphics[width=2.8in,height=2in]{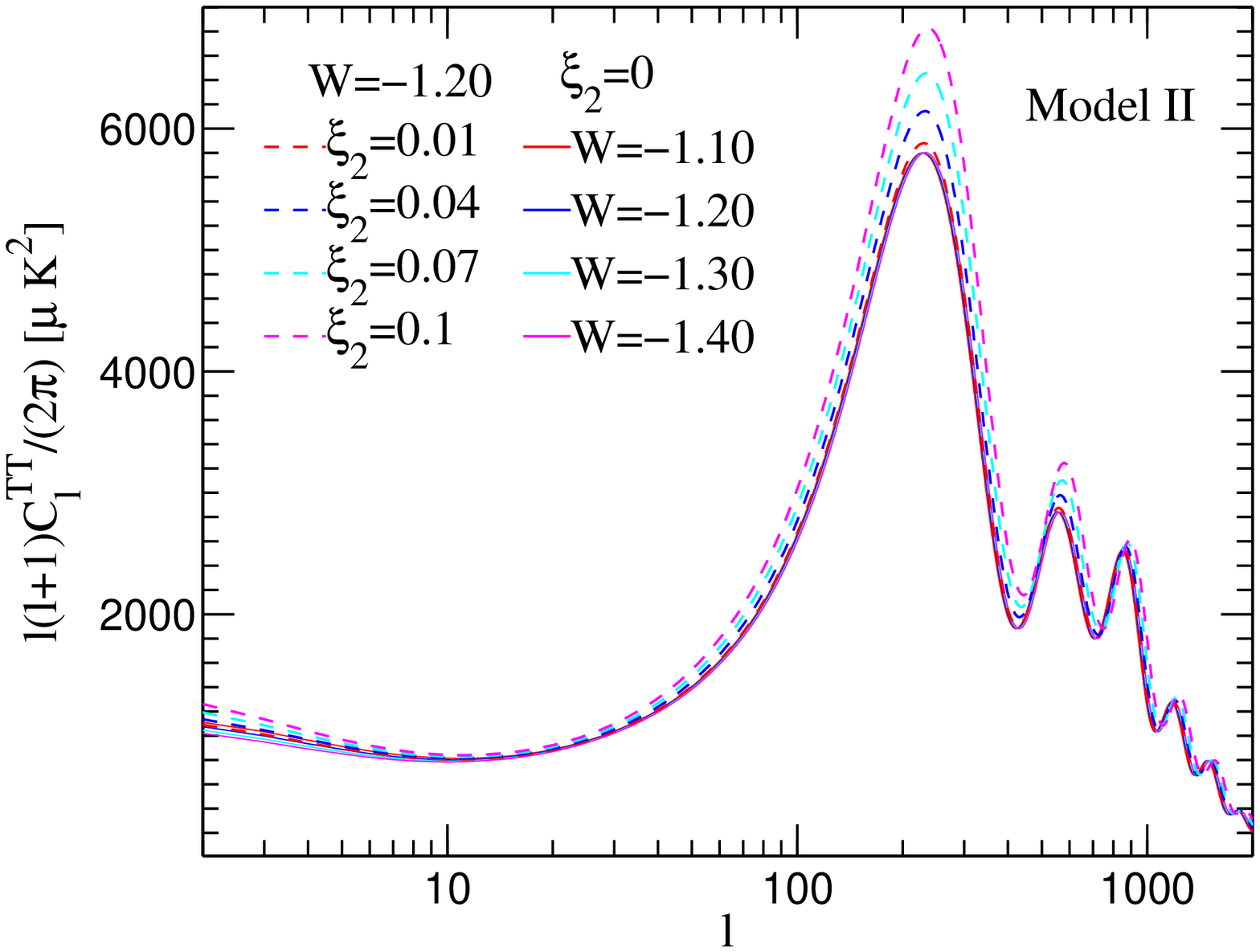}
\includegraphics[width=2.8in,height=2in]{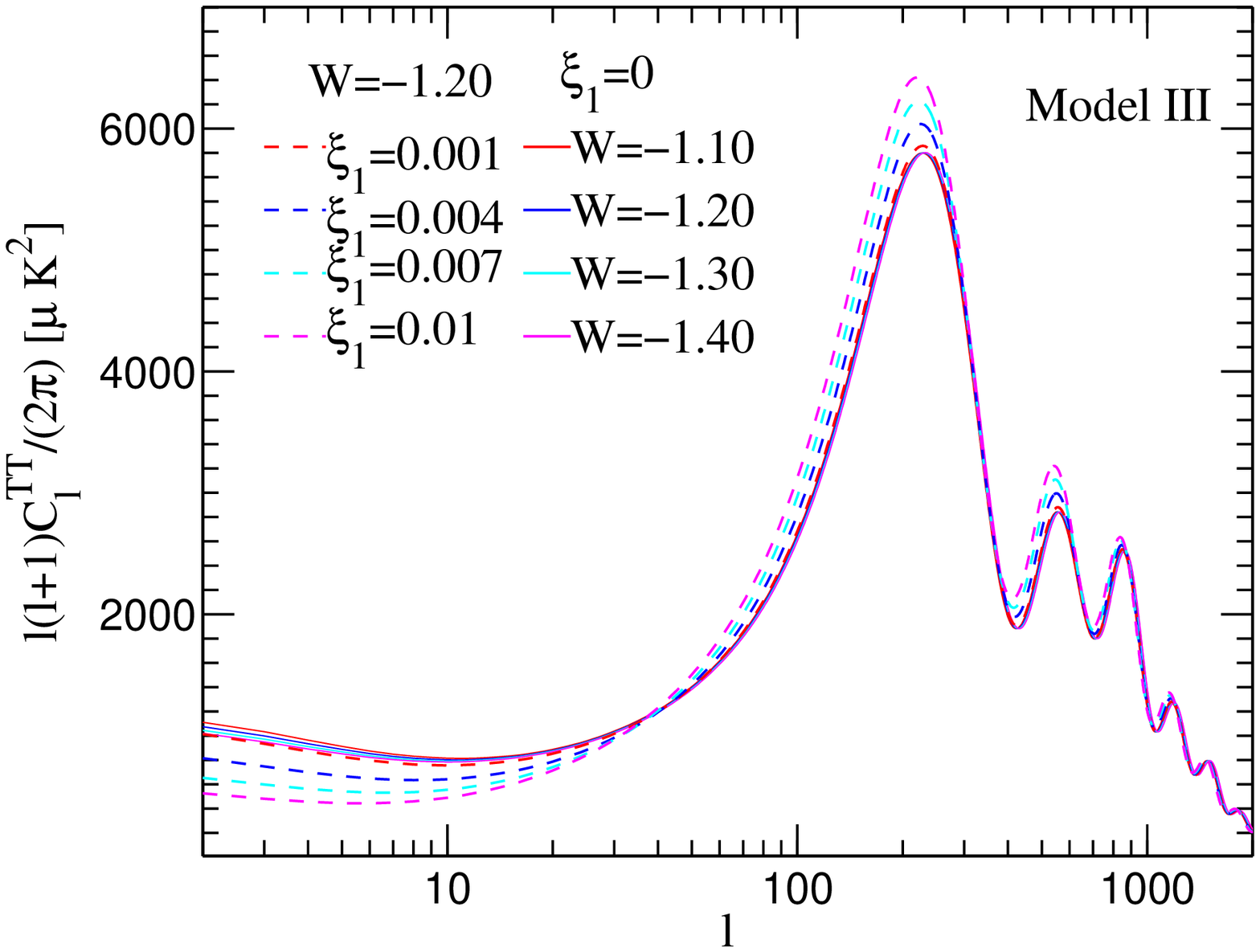}
\includegraphics[width=2.8in,height=2in]{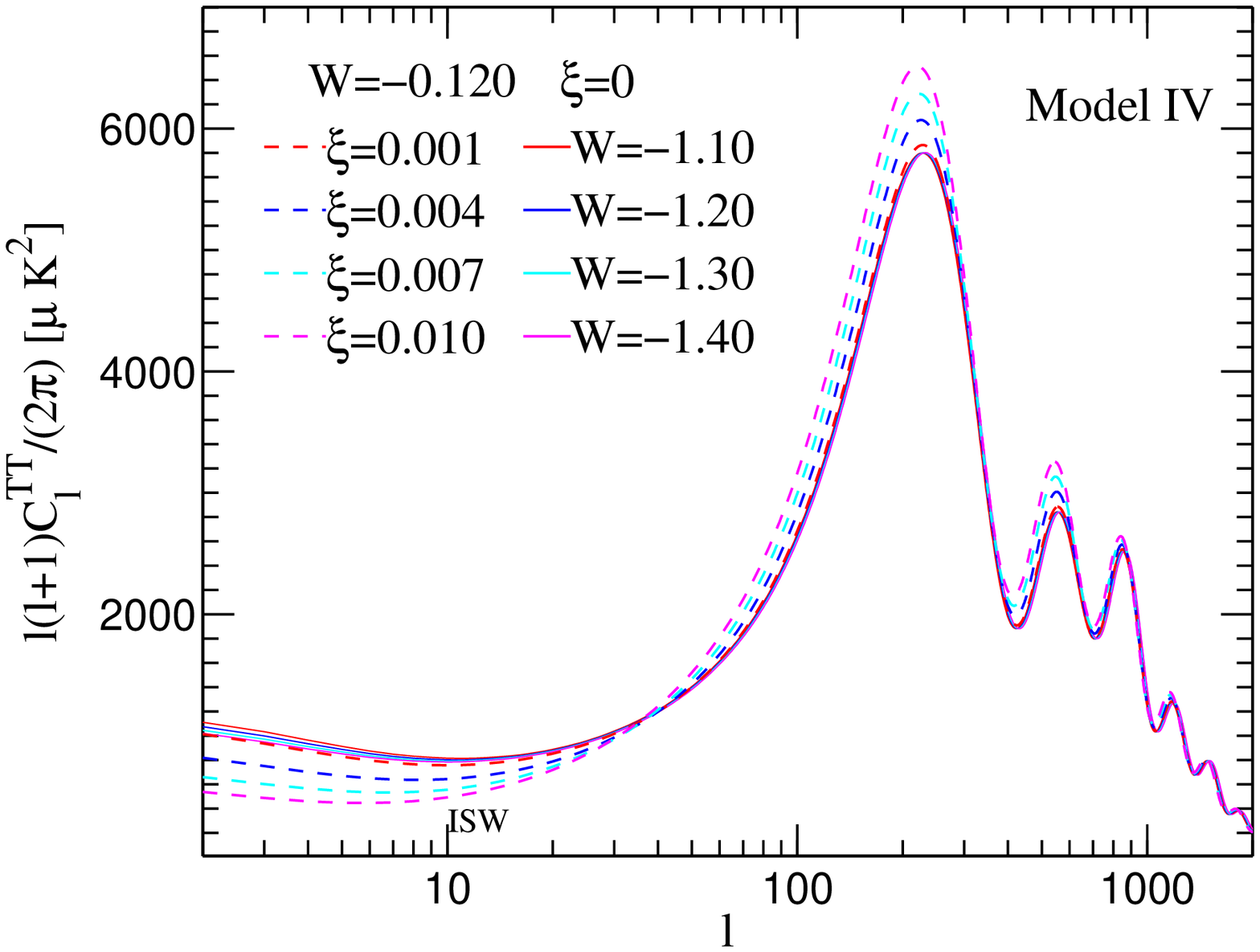}
\caption{The CMB TT power spectrum for different interaction models between DE and DM.}\label{cmb}
\end{figure*}

\begin{figure*}
\centering
\includegraphics[width=2.8in,height=2in]{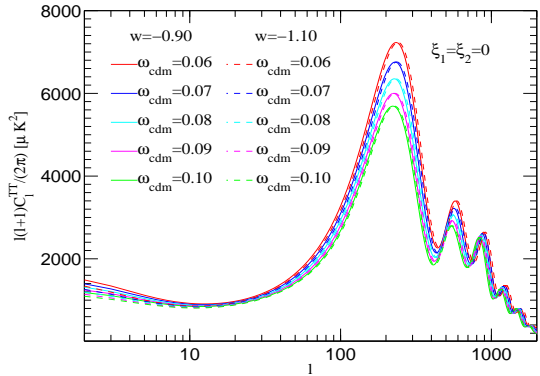}
\caption{The dependence of CMB angular power spectrum on physical cosmological parameter $\omega_c = \Omega_c h^2$}\label{omegac}
\end{figure*}

Now we are in a position to use the perturbation formalism to study the  influence of the interaction between dark sectors and other cosmological
parameters on the CMB power spectrum. In Figs.\ref{cmb} and \ref{omegac}  we illustrate the theoretical computation results of the CMB power
spectrum for different cosmological parameters.

Fixing the DM abundance, we see in the CMB TT angular power spectrum (Figs.\ref{cmb}) that the change of the constant DE EoS  only modifies the
low-l part of the spectrum while leaves the acoustic peaks almost unchanged. When the constant DE EoS $w>-1$, the low-l spectrum gets enhanced
with the increase of the value of $w$. Such a property keeps valid when the DE EoS parameter is a constant smaller than -1, namely the phantom
case $w<-1$. However when the DE is in the phantom region, the enhancement of the low-l spectrum due to the increase of the $w$ is less sensitive
than that of the quintessence DE.

In the low-l CMB spectrum, we see from Figs.\ref{cmb} that the coupling between dark sectors can also change the spectrum. As the coupling
becomes more positive, the low-l spectrum is further suppressed. When the interaction between dark sectors is proportional to the dark matter or
total dark sector energy density, the low-l spectrum is more sensitive to the change of the coupling than the DE EoS.

Beside the low-l CMB spectrum, the interaction between DE and DM can also influence the acoustic peaks. This feature is interesting, since this
property differs from that of the DE EoS and can help to break the degeneracy between the interaction between dark sectors and the DE EoS.

The above discussion is valid for fixed DM abundance. Now we investigate the dependence of CMB angular power spectrum on the abundance of cold
DM, $\omega_c=\Omega_ch^2$. Although the abundance of the DM does not affect much on the low-l CMB power spectrum, it quite influences the
amplitude of the first and second acoustic peaks in CMB TT angular power spectrum (see Fig. \ref{omegac}). Decreasing the cold DM abundance
$\omega_c$ will enhance the acoustic peaks. This effect is degenerated with the influence given by the dark sectors' interaction as we observed
in Fig.\ref{cmb}. A possible way to break this degeneracy is to consider the influence of the interaction on the low-l CMB spectrum. Moreover, we
can include further observations to get a complementary constraint on the DM abundance and this in turn can help to constrain the coupling
between dark sectors.

In order to extract the signature of the interaction and constraints on other cosmological parameters, we need to use the latest CMB data
together with other observational data. We perform such a task in the next section.

\section{global fitting and cosmological coincidence problem}
\begin{figure*}
\centering
\includegraphics[width=4.5in,height=4.5in]{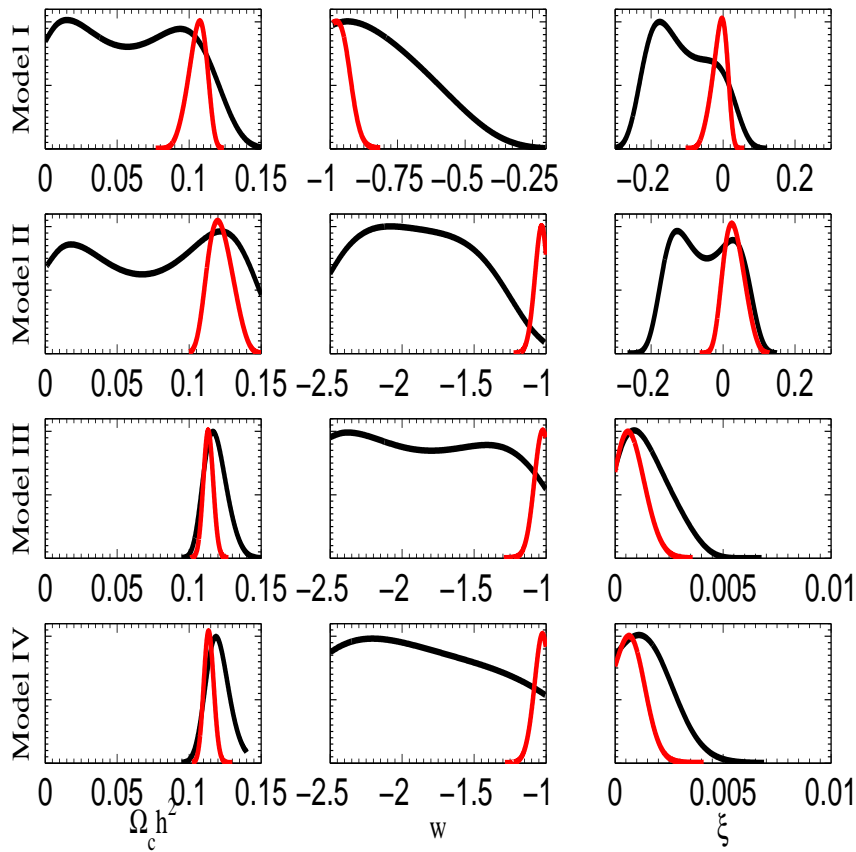}
\caption{The likelihood of cold dark matter abundance $\Omega_ch^2$, dark energy  EoS $w$ and couplings $\xi$ for different cosmological models.
The black lines denote the results obtained from WMAP 7 year data set alone and the red lines denote the results obtained from the combination of
$\rm{WMAP+SN+BAO+H_0}$}\label{wmapfig}
\end{figure*}

In this section we confront our models with observational
data by implementing joint likelihood analysis. We take the parameter space
as
$$P=(h,\omega_b,\omega_{cdm},\tau,\ln[10^{10}A_s],n_s,\xi_1,\xi_2,w)$$
where $h$ is the hubble constant, $\omega_b=\Omega_bh^2, \omega_{cdm}=\Omega_{cdm}h^2$, $A_s$ is the amplitude of the primordial curvature
perturbation, $n_s$ is the scalar spectral index, $\xi_1$ and $\xi_2$ are coupling constants proportional to the energy density of DM and DE
respectively, $w$ is the DE EoS. We choose the flat universe with $\Omega_k=0$ and our work is based on CMBEASY code\cite{easy}. We use the CMB
anisotropy data from the seven-year Wilkinson Microwave  Anisotropy Probe (WMAP). The Priors for Bayesian Analyses are presented in table
~\ref{Priors}. The fitting results in the $1\sigma$ range are listed in table ~\ref{summary}. We plot the likelihood for $\Omega_ch^2$, the
coupling between dark sectors and the DE EoS in Fig ~\ref{wmapfig}. It is clear that, when the interaction is proportional to the energy density
of DE, CMB data alone can not impose good constraints on $\Omega_ch^2$, the coupling and the DE EoS all together. This can be explained by our
theoretical analysis shown in Figs.\ref{cmb} and \ref{omegac}. At the low-l CMB spectrum, it is impossible to distinguish the DE EoS and the
coupling influences; in acoustic peaks it is hard to break the degeneracy between the coupling and the DM abundance.

When the interaction between dark sectors is proportional to the energy  density of DM or total dark sectors, CMB data alone can impose tight
constraints on couplings and $\Omega_ch^2$, but it can not impose good constraint on the DE EoS $w$. This result can also be understood from our
analysis in Figs.\ref{cmb} and \ref{omegac}. The degeneracy between the coupling and the DM abundance can be broken by looking at the low-l CMB
spectrum, since the CMB spectrum is not sensitive to the change of the DE EoS in the low-l spectrum. In order to get tighter constraint on
$\Omega_ch^2$, we use the BAO distance measurements \cite{BAO} which are obtained from analyzing clusters of galaxies and tests a different
region in the sky as compared to CMB. BAO measurements provide a robust constraint on the distance ratio
\begin{equation}
d_z = r_s(z_d)/D_v(z)
\end{equation}
where $D_v(z)\equiv[(1+z)^2D_A^2z/H(z)]^{1/3}$is the effective distance  \cite{Eisenstein}, $D_A$ is the angular diameter distance, and $H(z)$ is
the Hubble parameter. $r_s(z_d)$ is the comoving sound horizon at the baryon drag epoch where the baryons decoupled from photons. We numerically
find $z_d$ using the condition $\int_{\tau_d}^{\tau_0}\dot{\tau}/R=1,R=\frac{3}{4}\frac{\rho_b}{\rho_{\gamma}}$ as defined in \cite{wayne}. The
$\chi^2_{BAO}$ is calculated as \cite{BAO},
\begin{equation}
\chi^2_{BAO}=(\vec{\textbf{d}}-\vec{\textbf{d}}^{obs})^T\textbf{C}^{-1}(\vec{\textbf{d}}-\vec{\textbf{d}}^{obs})
\end{equation}
where $\vec{\textbf{d}}=(d_{z=0.2},d_{z=0.35})^T$, $\vec{\textbf{d}}^{obs}=(0.1905,0.1097)^T$ and the inverse of covariance matrix \cite{BAO}
\begin{equation}
\textbf{C}^{-1}=\left(
                  \begin{array}{cc}
                    30124 & -17227\\
                    -17227 & 86977 \\
                  \end{array}
                \right).
\end{equation}
Furthermore, we add the BAO A parameter \cite{BAO_A},
\begin{eqnarray}
A&=&\frac{\sqrt{\Omega_m}}{E(0.35)^{1/3}}\left[\frac{1}{0.35}\int_0^{0.35}\frac{dz}{E(z)}\right]^{2/3}\nonumber \\
 &=&0.469(n_s/0.98)^{-0.35}\pm 0.017
\end{eqnarray}
where $E(z)=\frac{H(z)}{H_0}$ and $n_s$ are the scalar spectral index. In order to improve the constraints on the DE EoS $w$, we use the
compilation of  397 Constitution samples from supernovae survey \cite{SNeIa}. We compute
\begin{equation}
\chi^2_{SN}=\sum \frac{[\mu(z_i)-\mu_{obs}(z_i)]^2}{\sigma_i^2}\quad ,
\end{equation}
and marginalize the nuisance parameter.
In addition to the above mentioned data sets, we also add the latest constraint on the present-day Hubble constant \cite{Hubble},
\begin{equation}
H_0=74.2\pm 3.6 {\rm km  s^{-1} Mpc^{-1}}.
\end{equation}
We implement the joint likelihood analysis that is,
\begin{equation}
\chi^2=\chi^2_{WMAP}+\chi^2_{SN}+\chi^2_{BAO}+\chi^2_{H_0}.
\end{equation}
The fitting  results are shown in Table \ref{full}. The cosmological parameters are well constrained. When the coupling between dark sectors is
proportional to the energy density of DE, its value is constrained up to a few percent. When the coupling is proportional to the energy density
of DM or total dark sectors, its constraint is pretty good and reads $\xi_1=0.0006^{+0.0006}_{-0.0005}$ and $\xi= 0.0006^{+0.0005}_ {-0.0006}$
respectively. In 1 $\sigma$ range the couplings are positive. The likelihoods of the fitting results for the DM abundance, DE EoS and the
coupling between dark sectors are shown in Fig ~\ref{wmapfig}. Compared with the WMAP data alone, we see that the joint analysis by including
other observational data provides tighter constraints on the cosmological parameters.

\begin{figure}
\includegraphics[width=3in,height=2.8in]{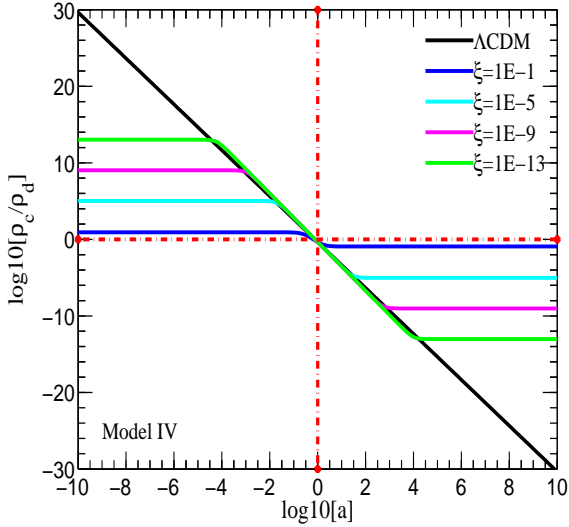}
\caption{Cosmological coincidence problem}\label{coincidence}
\end{figure}

\begin{figure}
\includegraphics[width=3in,height=2.8in]{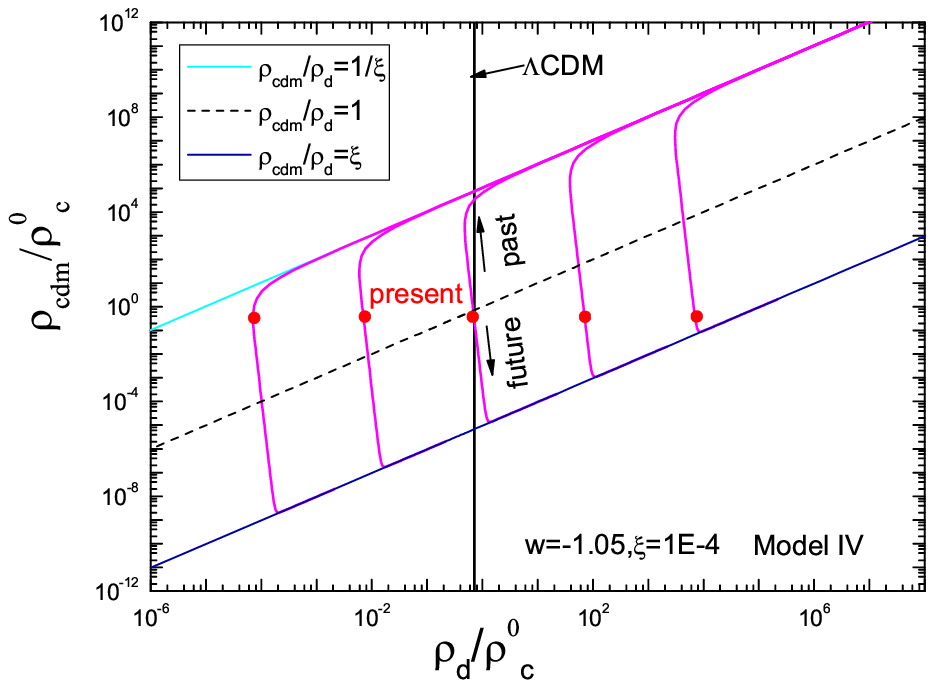}
\caption{$\rho_c^0$ is the critical energy density today. The attractor solutions  of $r$ does not depend on the initial conditions at the early
time of the universe. The purple lines represent the density evolution of cosmological model with different initial conditions. Noted by points,
the density contrast $r$ today are different for different initial conditions but they are bounded in two  attractor solutions
$r_1\sim\xi,r_2\sim1/\xi$ in $\rho_c-\rho_d$ plane. }\label{coinci}
\end{figure}

The positive coupling can help us to alleviate the cosmological coincidence problem\cite{hePRD09,heJCAP08}. As shown in Fig~\ref{coincidence},
the energy density of DE and DM in standard $\Lambda{\rm CDM}$ model are only comparable at present moment. The thick black line representing the
quantity $\log_{10}(\rho_c/\rho_d)$ is linearly proportional to $\log_{10}a$ and almost precisely crosses the origin in the whole expansion
history of the universe. However, this is hard to be convincing and achieving. If we want the energy density of DM to be comparable to that of
the DE at the present moment, such an origin crossing can only be realized by tuning the initial conditions at the early time of the universe
over $30$ orders in energy density contrast $r=\rho_c/\rho_d$.  If there is little change in the initial condition, the
$\log_{10}(\rho_c/\rho_d)$ cannot cross the origin and at the present the energy densities of DM and DE cannot be comparable. This problem can be
overcome by introducing the interaction between DE and DM. As an example we show the model when the interaction between DE and DM is proportional
to the energy density of the total dark sectors (Model IV).  In this case there are two attractor solutions of the ratio $r$ during the expansion
history of the universe
\begin{eqnarray}
r_1\sim\frac{1}{\xi}\nonumber \\
r_2\sim\xi,
\end{eqnarray}
by considering $\xi$ is a small value from the fitting results. $r_1$  happened  in the past and $r_2$ will occur in the future. The behavior of
the attractor solutions of the ratio $r$ only depends on the coupling constant $\xi$ and does not depend on the initial conditions at the early
time of the universe. To see this point more clearly, we show that in Fig~\ref{coinci}, the purple lines represent the density evolution of
cosmological model with different initial conditions. The density contrast $r$ at present is different for different initial conditions. However,
they are all bounded in two attractor solutions $r_1\sim\xi,r_2\sim1/\xi$ in the plane. Adopting the coupling constant value from the fitting,
$1/\xi\sim 10^4$ and $\xi \sim 10^{-4}$, we have the ratio $r$  in the range $10^{-4}<r<10^{4}$ during the universe history. Thus the change of
the ratio $r$ is much smaller than that of the $\Lambda{\rm CDM}$ model so that the period when the DE and DM are comparable is much longer than
that of the $\Lambda{\rm CDM}$ model. The cosmological coincidence problem can thus be greatly alleviated.
\begin{table*}
\caption{Priors for Bayesian Analyses }\label{Priors}
\begin{tabular}{c}
\hline
\hline
$0<\Omega_ch^2<0.5$  \\
$-1<w<-0.1$(Model I)\quad $-2.5<w<-1$(Model II,III,IV)  \\
$-0.4<\xi<0.4$ (Model I,II)\quad $0<\xi<0.02$(Model,III,IV) \\
\hline
\end{tabular}
\end{table*}

\begin{table*}
\caption{The fitting results using WMAP 7-year data alone }\label{summary}
\begin{tabular}{|c||c|c|c|c|}
\hline
\hline
Parameters & Model I & Model II & Model III & Model IV\\
\hline
$h$ & $0.678^{+0.061}_{-0.075}$ &$1.09^{+0.23}_{-0.26}$ &$0.80^{+0.21}_{-0.13}$ & $0.83^{+0.36}_{-0.15}$ \\
\hline
$\Omega_bh^2$ &$0.0224^{+0.0006}_{-0.0006}$ &$0.0221^{+0.0005}_{-0.0005}$ &$0.0219^{+0.0006}_{-0.0006}$ & $0.0219^{+0.0006}_{-0.0006}$ \\
\hline
$\Omega_ch^2$ &$<0.111$(68\%{\rm CL}) &$<0.151$(68\%{\rm CL}) & $0.117^{+0.009}_{-0.007}$ & $0.119^{+0.008}_{-0.007}$ \\
\hline
$\tau$ & $0.084^{+0.015}_{-0.014}$ & $0.085^{+0.015}_{-0.014}$ & $0.087^{+0.016}_{-0.015}$& $0.085^{+0.016}_{-0.015}$ \\
\hline
$n_s$ &$ 0.966^{+0.014}_{-0.015}$ & $0.957^{+0.014}_{-0.014}$ & $0.944^{+0.016}_{-0.016}$ & $0.943^{+0.017}_{-0.018}$ \\
\hline
$\rm{ln}[10^{10}As]$ & $3.071^{+0.037}_{-0.036}$ &$3.072^{+0.036}_{-0.035}$ &$3.079^{+0.039}_{-0.038}$ & $3.077^{+0.037}_{-0.035}$ \\
\hline
$w$ & $<-0.694$(68\%{\rm CL}) & unconstrained &unconstrained & unconstrained \\
\hline
$\xi$ & $-0.17^{+0.17}_{-0.05}$ & $ -0.13^{+0.20}_{-0.05}$ & $0.0010^{+0.0012}_{-0.0010}$& $0.0011^{+0.0010}_{-0.0011}$ \\
\hline
\hline
\end{tabular}
\end{table*}
\begin{table*}
\caption{The fitting results using $\rm{WMAP+SN+BAO+H_0}$\label{full}}
\begin{tabular}{|c||c|c|c|c|}
\hline
\hline
Parameters & Model I & Model II & Model III & Model IV\\
\hline
$h$ & $0.699^{+0.012}_{-0.012}$ &$0.709^{+0.013}_{-0.012}$ &$0.700^{+0.013}_{-0.013}$ & $0.699^{+0.013}_{-0.013}$ \\
\hline
$\Omega_bh^2$ &$0.0224^{+0.0006}_{-0.0006}$ &$0.0222^{+0.0005}_{-0.0005}$ &$0.0222^{+0.0006}_{-0.0006}$ & $0.0222^{+0.0006}_{-0.0006}$ \\
\hline
$\Omega_ch^2$ &$0.107^{+0.006}_{-0.007}$ &$0.120^{+0.010}_{-0.008}$ & $0.113^{+0.003}_{-0.003}$ & $0.114^{+0.003}_{-0.003}$ \\
\hline
$\tau$ & $0.086^{+0.016}_{-0.015}$ & $0.083^{+0.016}_{-0.014}$ & $0.087^{+0.017}_{-0.015}$& $0.087^{+0.016}_{-0.015}$ \\
\hline
$n_s$ &$ 0.967^{+0.013}_{-0.013}$ & $0.961^{+0.013}_{-0.013}$ & $0.956^{+0.014}_{-0.014}$ & $0.956^{+0.014}_{-0.014}$ \\
\hline
$\rm{ln}[10^{10}As]$ & $3.070^{+0.036}_{-0.034}$ &$3.069^{+0.035}_{-0.033}$ &$3.074^{+0.038}_{-0.036}$ & $3.074^{+0.036}_{-0.034}$ \\
\hline
$w$ & $<-0.938$(68\%{\rm CL}) & $-1.03^{+0.03}_{-0.04}$ &$-1.02^{+0.02}_{-0.05}$ & $-1.03^{+0.03}_{-0.05}$ \\
\hline
$\xi$ & $-0.003^{+0.017}_{-0.024}$ & $0.024^{+0.034}_{-0.027}$ & $0.0006^{+0.0006}_{-0.0005}$& $0.0006^{+0.0005}_{-0.0006}$ \\
\hline
\hline
\end{tabular}
\end{table*}
\section{conclusions and discussions}
In this paper we have reviewed the formalism of the perturbation theory when there  is an interaction between DE and DM. We have proposed a way
to construct the coupling vector in a self consistent manner both in the perturbed form and in the background. Based upon the perturbation
formalism we have studied the signature of the interaction between dark sectors from CMB angular power spectrum. Theoretically we found that
there are possible ways to break the degeneracy between the interaction, DE EoS and DM abundance. This can help to get tight constraint on the
interaction between DE and DM.

We have performed the global fitting by using the CMB power spectrum data from WMAP7Y results together with latest SNIa, BAO and $\rm{H_0}$ data
to constrain the interaction between DE and DM. When the interaction between DE and DM takes the form proportional to the energy density of DM
and the total dark sectors, in $1\sigma$ range the coupling is found to be positive. The tight positive coupling indicates that there is energy
flow from DE to DM, which can help us to alleviate the cosmological coincidence problem.

The question of how to improve the model is now much related to find a field theory based model for the interaction and how to relate the
model to the standard model of particle interactions. This is currently under study.

\textbf{Acknowledgement:} This work has been supported partially by NNSF of China No. 10878001 and the National Basic Research Program of China
under grant 2010CB833000. EA wishes to thank FAPESP and CNPq (Brazil) for financial support.

\section{Appendix: }
\subsection{covariant equation of motion for interacting system\label{mes}}
The basic dynamics of interacting systems in classical mechanics is described by Meshchersky's equation,
\begin{equation}
m\frac{dv^a}{dt}=\frac{dm}{dt}(v_t^a-v^a)+F^a,\label{new}
\end{equation}
where $\frac{dm}{dt}$ is the rest mass(energy) transfer rate by the moving system,  $F^a$ is  the external force, $\frac{dm}{dt}v_t^a$ is the
momentum transfer and $\frac{dm}{dt}v^a$ is the inertial force.

The classical Meshchersky's equation can be extended  to  special relativity and general relativity. In the framework of special relativity, it
was first derived by Ackert \cite{Ackeret} and then summarized by Seifert in \cite{summerfield}. Here we first extend their results on the
equations of motion to a covariant form, then generalize them to curved spacetime. We consider a moving system with rest energy density $\rho$
and the energy-momentum tensor,
\begin{equation}
T^{ab}=\rho U^aU^b
\end{equation}
where $U^a=(\frac{\partial}{\partial \tau})^a$ is the four velocity.
The covariant form of the equation of motion can be given by,
\begin{equation}
\partial_bT^{ab}=QU_t^a+Q_F^a\label{em}
\end{equation}
where $U_t^a$ is the energy transfer four velocity and $Q_F^a$ is the external four  force density acting on the system. For a given inertial
observer $Z^a=(\frac{\partial}{\partial t})^a$ in Minkowski space-time, the ordinary derivative operator $\partial_b$ vanishes on $Z^a$,
$\partial_b Z^a=0$. The time-like part of the above equation reads,
\begin{eqnarray}
Z_a\partial_bT^{ab}&=&\dot{\rho} Z_aU^a+\rho Z_aA^a+Z_aU^a\rho\Theta\nonumber \\
                 &=&-\frac{d}{d\tau}(\gamma \rho)-\gamma\rho \Theta\nonumber\\
                 &=&-\gamma_tQ+Z_aQ_F^a\label{energyM}
\end{eqnarray}
where $A^a=U^b\partial_bU^a$ is the four acceleration, dot denotes  $U^b\partial_b=\frac{\partial}{\partial \tau}$,$\Theta=\partial_aU^a$,
$\gamma=-Z_aU^a,\gamma_t=-Z_aU_t^a$ are Lorentz-boost factors and $Z_aQ_F^a$ represents the energy transfer density observed by $Z^a$.

The space-like part reads,
\begin{eqnarray}
{h_a}^c\partial_bT^{ab}&=&\dot{\rho}{h_a}^cU^a+\rho{h_a}^cA^a+\rho{h_a}^cU^a\Theta\nonumber \\
                     &=&\frac{d}{d\tau}(\gamma \rho v^c)+\rho\gamma v^c\Theta\nonumber\\
                     &=&Q\gamma_tv_t^c+{h_a}^cQ_F^a\label{momentumM}
\end{eqnarray}
where ${h_a}^c={\delta_a}^c+Z_aZ^c$ is the projection operator, $v^c={h_a}^cU^a/\gamma,v_t^c={h_a}^cU_t^a/\gamma_t$ are three velocities observed
by $Z^a$ and ${h_a}^cQ_F^a$ is the three force density acting on the moving system. If there is no expansion $\Theta=0$ in the system,
Eqs.~(\ref{energyM}) and ~(\ref{momentumM}) go back to Eqs.(125) and (126) in \cite{summerfield}.

With the help of the covariant form, Eq.~(\ref{em}) can be generalized to curved spacetime  by the ``minimal substitution" $\partial_b\rightarrow
\nabla_b$,
\begin{equation}
\nabla_bT^{ab}=QU_t^a+Q_F^a=Q^a\label{ge}
\end{equation}
The above equation is the generalized Meshchersky's equation which is  the basic equation of motion for interacting systems in curved spacetime.

In order to give a clear physical interpretation on this equation, we study the dynamics in terms of the distribution function. The
energy-momentum tensor can be written as \cite{Christos},
\begin{equation}
T^{ab}=\int fp^ap^b\frac{d^3\vec{p}}{m}\label{bolzman}
\end{equation}
where $f=f(x^a, p^a)$ is a distribution function, $m$ is the rest mass  for moving particles, $p^a=mU^a$ is the four momentum and
$\frac{d^3\vec{p}}{m}$ is the Lorentz-invariant volume element on the positive-energy mass shell. Eq. ~(\ref{ge}) can be presented in the form
\begin{equation}
\nabla_bT^{ab}=\int C[f]p^a\frac{d^3\vec{p}}{m}+\int m\frac{df_*}{ds}p^a\frac{d^3\vec{p}}{m}.
\end{equation}
If contracted with the given four velocity, the LHS of the above equation  gives rise to the monopole and dipole of Boltzmann equations.  The
coupling vector $Q^a$ on the RHS consists of two terms but with distinct physical meanings. The first term is the four force density produced by
collisions, which is a space-like vector.
\begin{equation}
Q_F^a=\int C[f]p^a\frac{d^3\vec{p}}{m}\nonumber
\end{equation}
where $C[f]$ is the collision kernel. The second term is the energy momentum transfer density along the  direction of the average four velocity
$U_t^a=\frac{\partial}{\partial s}$. $QU_t^a$ is a time-like vector.
\begin{equation}
QU_t^a=\int m\frac{df_*}{ds}p^a\frac{d^3\vec{p}}{m}\label{qt}
\end{equation}
$\frac{df_*}{ds}$ refers to the change rate in distribution function  due to the varying rest mass of particles or varying comoving particle
number density in the system investigated. However, it needs to be specified according to physics. If decomposed relative to a given observer
with four velocity $Z^a$, $p^a$ can be represented as
\begin{equation}
p^a=m\gamma(\vec{p})(Z^a+v^a(\vec{p}))
\end{equation}
where $\gamma(\vec{p})=-p_aZ^a/m$. Therefore, the time-like part of Eq.~(\ref{qt}) reads
\begin{equation}
\gamma Q=\int m\frac{df_*}{ds}\gamma(\vec{p})d^3\vec{p}
\end{equation}
where $\gamma = -U_t^aZ_a$ and $\gamma Q$ represent the energy transfer rate observed by $Z^a$. The spatial part reads
\begin{equation}
\gamma Q v_t^a=\int m\frac{df_*}{ds}\gamma(\vec{p})v^a(\vec{p})d^3\vec{p}
\end{equation}
where $v_t^a$ is the average energy transfer velocity,
\begin{equation}
v_t^a=\frac{\int m\frac{df_*}{ds}\gamma(\vec{p})v^a(\vec{p})d^3\vec{p}}{\int m\frac{df_*}{ds}\gamma(\vec{p})d^3\vec{p}}\label{velocit}
\end{equation}
In hydrodynamics, $\gamma Q v_t^a$ refers to the viscosity due to the momentum transfer in different components.

The above equations are quite general and now we concentrate our  discussion on the DE and DM coupled system. The external non gravitational
force density acting on the system vanishes $Q_F^a=0$ in the background due to the homogeneous universe.   Furthermore, noting that the spacetime
is isotropic, $\frac{df_*}{ds}$ only depends on time and the contribution of $v^a$ in Eq. ~(\ref{velocit}) is counteracted in opposite
directions. $v_t^a$ vanishes in the background. Only the energy transfer $Q$ can be observed in the background.

In the perturbed universe, as we have neglected the anisotropic stress-tensor, we assume that the perturbation $\frac{d\delta f_*}{ds}$ is still
isotropic, from Eq.~(\ref{velocit}) we find that $v_t^a$ vanishes and the coupling vector is independent of the bulk motion of the component in
the universe.
\subsection{covariant coupling vector in perturbed spacetime\label{appendix}}
The general coupling vector $ {Q_{(\lambda)}}^{\nu}$ is independent of  the choice of observer. But in the literatures, it is usually decomposed
in two parts with respect to a given observer $\lambda'$ with four velocity ${U_{(\lambda')}}^{\mu}$.
\begin{eqnarray}
{Q_{(\lambda)}}^{\mu}&=&Q{_{(\lambda \lambda')}}{U_{(\lambda')}}^{\mu}+{F_{(\lambda \lambda')}}^{\mu}\nonumber \\
&=&-({U_{(\lambda')}}_{\nu}Q{_{(\lambda)}}^{\nu}){U_{(\lambda')}}^{\mu}+{{h_{(\lambda')}}^{\mu}}_{\nu}{Q_{(\lambda)}}^{\nu} \label{couplevector}
\end{eqnarray}
where $Q{_{(\lambda  \lambda')}}=-{U_{(\lambda')}}_{\nu}Q{_{(\lambda)}}^{\nu}$ is the energy transfer rate of $\lambda$ component observed by the
$\lambda'$ observer and ${F_{(\lambda \lambda')}}^{\mu}={{h_{(\lambda')}}^{\mu}}_{\nu}{Q_{(\lambda)}}^{\nu}$ is the momentum transfer observed by
the $\lambda'$ observer, correspondingly.

We can show that such a decomposition cannot bring substantial physics  because Eq.~(\ref{couplevector}) is an identity. Furthermore, we can show
that the perturbed forms are also identities. The perturbation of the zero-th component on the RHS of Eq.~(\ref{couplevector}) reads
\begin{eqnarray}
&&(\delta Q{_{(\lambda \lambda')}}){U_{(\lambda')}}^0 + (Q{_{(\lambda \lambda')}})\delta{U_{(\lambda')}}^0+\delta{F_{(\lambda \lambda')}}^0 \nonumber \\
&&=-(\delta {U_{(\lambda')}}_{\nu})Q{_{(\lambda)}}^{\nu}{U_{(\lambda')}}^0-  ({U_{(\lambda')}}_{\nu}\delta Q{_{(\lambda)}}^{\nu}){U_{(\lambda')}}^0 \nonumber \\
&&- ({U_{(\lambda')}}_{\nu}Q{_{(\lambda)}}^{\nu})\delta{U_{(\lambda')}}^0+ {\delta {h_{(\lambda')}}^0}_{\nu}{Q_{(\lambda)}}^{\nu}\nonumber \\
&&+{{h_{(\lambda')}}^0}_{\nu}\delta{Q_{(\lambda)}}^{\nu}=\delta {Q_{(\lambda)}}^0
\end{eqnarray}
where we have used $U_{0(\lambda')}=-a$, ${U_{(\lambda')}}^0=1/a$, $\delta {U_{(\lambda')}}^0 = -\psi {U_{(\lambda')}}^0$, $\delta
U_{{(\lambda')}0} =\psi U_{{(\lambda')}0}$ in the derivation. We find that the nonzero $\delta{F_{(\lambda \lambda')}}^0$ plays an
important role in getting the identity in zero-th component.

Similarly, the perturbation of the spatial component in the RHS of Eq.~(\ref{couplevector}) reads
\begin{eqnarray}
&&(\delta Q{_{(\lambda \lambda')}}){U_{(\lambda')}}^i+Q{_{(\lambda \lambda')}}\delta {U_{(\lambda')}}^i+\delta {F_{(\lambda \lambda')}}^i \nonumber \\
&&=-\delta ({U_{(\lambda')}}_{\nu}{Q_{(\lambda)}}^{\nu}){U_{(\lambda')}}^i-({U_{(\lambda')}}_{\nu}{Q_{\lambda}}^{\nu})\delta {U_{(\lambda')}}^i\nonumber \\
&&+\delta({{h_{(\lambda')}}^i}_{\nu})Q_{(\lambda)}^{\nu}+{{h_{(\lambda')}}^i}_{\nu}\delta Q_{(\lambda)}^{\nu}\nonumber \\
&&=-({U_{(\lambda')}}_0{Q_{(\lambda)}}^0)\delta {U_{(\lambda')}}^i+({U_{(\lambda')}}_0{Q_{\lambda}}^0)\delta {U_{(\lambda')}}^i\nonumber\\
&&+{{h_{(\lambda')}}^i}_{\nu}\delta {Q_{(\lambda)}}^{\nu}=
\delta {Q_{(\lambda)}}^i
\end{eqnarray}
where ${U_{(\lambda')}}^i=0$, ${Q_{(\lambda')}}^i=0$, $\delta {U_{(\lambda')}}^i=\partial^i v_{(\lambda')}/a$ have been employed.  The net effect
is ${{h_{(\lambda')}}^i}_b\delta {Q_{(\lambda)}}^b=\delta {Q_{(\lambda)}}^i$. Thus the i-th component is also an identity.

Since ${Q_{(\lambda)}}^{\nu}$ does not depend on observer, we need to  specify such a coupling vector directly as discussed in \ref{mes}. Once it
is specified in the background, the zero-th component of the perturbed form $\delta Q_{(\lambda)}^0$ can be uniquely determined by the background
$Q_{(\lambda)}^{\mu}$. For this purpose, we consider the module of $Q_{(\lambda)}^{\mu}$
\begin{equation}
{Q_{(\lambda)}}^{\mu}Q_{(\lambda)\mu}=g_{00}({Q_{(\lambda)}}^0)^2=-{Q_{(\lambda)}}^2,
\end{equation}
where $Q_{(\lambda)}=a{Q_{(\lambda)}}^{0}$ is a scalar on FRW space and  the minus sign comes here because  ${Q_{(\lambda)}}^{\mu}$ is time-like.
By considering the perturbation of the above equation, we find
\begin{equation}
-\frac{\psi}{a} Q_{(\lambda)} + \frac{1}{a}\delta Q_{(\lambda)}=\delta Q_{(\lambda)}^0
\end{equation}
where $\psi$ arises from $\delta g_{00}$. The first term is from the  perturbation  $\delta e_0=-\psi/a$  and the second term comes from the
perturbation of the module.

$Q_{(\lambda)}$ is a scalar and under gauge transformation
\begin{equation}
\delta \tilde{Q}_{(\lambda)} =\delta Q_{(\lambda)} - \mathcal{L}_{\xi^0}Q_{(\lambda)}=\delta Q_{(\lambda)} - Q'_{(\lambda)}\xi^0.\label{lie}
\end{equation}
Noting \cite{31},
\begin{equation}
\tilde{\psi} = \psi -\xi^{0'}-\frac{a'}{a}\xi^0
\end{equation}
we find
\begin{eqnarray}
\delta \tilde{Q}_{(\lambda)}^0&=&-\psi Q_{(\lambda)}^0 + \frac{Q_{(\lambda)}}{a^2 Q_{(\lambda)}^0}\delta Q_{(\lambda)}-Q_{(\lambda)}^{0'}\xi^0+Q_{(\lambda)}^0\xi^{0'}\nonumber \\
& = & \delta Q_{(\lambda)}^0 -Q_{(\lambda)}^{0'}\xi^0+Q_{(\lambda)}^0\xi^{0'}
\end{eqnarray}
which is consistent with the gauge transformation of  $\delta Q_{(\lambda)}^0$ required by a covariant vector ${Q_{(\lambda)}}^{\mu}$\cite{31}.

The spatial part  $\delta Q^i_{(\lambda)}$  is independent of the zero-th  component. It refers to non-gravitational force. The covariant
perturbation of the potential $\delta Q_{p\lambda}$ of the spacial part $\delta Q^i_{(\lambda)}$ can be written as,
\begin{eqnarray}
\delta Q_{p\lambda}=\delta Q_{p\lambda}^I\mid_{\lambda'}+Q_{(\lambda)}^0v_{(\lambda')}\label{deltaQ}
\end{eqnarray}
where $\delta Q_{p\lambda}^I\mid_{\lambda'}$ is the perturbation $\delta Q_{p\lambda}$  observed in $\lambda'$ rest frame and
$Q_{(\lambda)}^0v_{(\lambda')}$ represents doppler effect, $\delta Q_{p\lambda}$ satisfies the gauge transformation
\begin{equation}
\delta \tilde{Q}_{p\lambda}=\delta Q_{p\lambda}+Q_{(\lambda)}^0\beta'.
\end{equation}
In DE and DM coupled system, we assume that $\delta Q_{p\lambda}^I\mid_{\lambda'}$  vanishes in the background $\delta
Q_{p\lambda}^I\mid_{back}=0$, where $v_{(back)}=0$, and
\begin{eqnarray}
\delta Q_{p\lambda}=\delta Q_{p\lambda}^I\mid_{back}+Q_{(\lambda)}^0v_{(back)}=0\label{deltaQ}
\end{eqnarray}
There are no non-gravitational force and doppler effect produced by energy transfer.
\subsection{gauge conditions\label{gauge}}
\subsubsection{Conformal Newtonian gauge}
The conformal newtonian gauge is a set of coordinates in which the perturbed line element satisfies
\begin{equation}
B=E=E'=0\quad.
\end{equation}
The gauge is fully fixed and thus the Barddeen's potential \cite{Bardeen} can be simply calculated as
\begin{eqnarray}
\Psi = \psi \quad,\nonumber \\
\Phi = \phi\quad. \nonumber
\end{eqnarray}
The gauge condition fixes the expressions for the gauge transformation
\begin{eqnarray}
\xi^0&=&-\frac{\tilde{B}}{k}-\frac{\tilde{E}'}{2k^2}\nonumber \\
\beta&=&\frac{\tilde{E}}{2k}\quad.
\end{eqnarray}
In particular, when we take the gauge transformation in different Conformal Newtonian coordinates $ B=\tilde{B}=E=\tilde{E}=E'=\tilde{E}'=0$,
$\xi^0=\beta=0$, all the perturbations will have the same value eg. $\tilde{\delta}_{(\lambda)}= \delta_{(\lambda)}$,  so that the Conformal
Newtonian gauge yields unambiguous results.
\subsubsection{Gauge mode in Synchronous gauge and weak equivalence principle}
Synchronous gauge is defined by $\psi=B=0$. The gauge invariant Barddeen's potential \cite{Bardeen} in Synchronous gauge can be calculated by,
\begin{eqnarray}
\Psi&=&\frac{\mathcal{H}}{2k^2}(6\eta_T+h_L)'+\frac{1}{2k^2}(6\eta_T+h_L)''\nonumber \\
\Phi&=&-\eta_T+\frac{\mathcal{H}}{2k^2}(6\eta_T+h_L)'\quad ,
\end{eqnarray}
where $\eta_T$, $h_L$ are Synchronous gauge parameters,
\begin{eqnarray}
\eta_T&=&-(\phi+\frac{E}{6})\nonumber \\
h_T &=& 6\phi \quad.
\end{eqnarray}

In contrast to Conformal Newtonian gauge, the metric conditions do not fully specify the gauge and need to be  supplemented by additional
definitions. When taking the gauge transformation in Synchronous coordinates $\psi=\tilde{\psi}=B=\tilde{B}=0$, it defines the gauge
transformation up to two arbitrary constants $ C_1$, $C_2$. These constants manifest themselves in time and spatial coordinate transformation
\cite{Sasaki}
\begin{eqnarray}
\xi^0&=&C_1a^{-1}\nonumber \\
\beta&=&-kC_1\int\frac{d\tau}{a} + C_2\quad .
\end{eqnarray}
The ambiguity of $C_1$ and $C_2$ leads to the gauge modes in density and velocity perturbations \cite{Sasaki},
\begin{eqnarray}
\tilde{\delta}_{\lambda}\mid_s&=&\delta_{\lambda}\mid_s -\frac{\rho_{\lambda}'}{\rho_{\lambda}}\frac{C_1}{a}\nonumber \\
\tilde{v}_{\lambda}\mid_s&=&v_{\lambda}\mid_s-k\frac{C_1}{a}\quad ,
\end{eqnarray}
where $s$ indicates that the perturbations are confined on different Synchronous coordinates.  The condition  $\psi=B=0$ does not yield
unambiguous results and additional definitions are called for.

Usually, $C_2$ can be obtained by fixing the initial curvature perturbation \cite{Sasaki}, while one gets $C_1$ fixing the peculiar velocity of
free falling non relativistic species in the universe. It is usual to set the peculiar velocity of the cold DM to be zero $v_c=0$ throughout the
expansion history,
$$\tilde{v}_{c}=v_{c}=C_1=0\quad .$$
Hence the condition $\psi=B=v_{c}=0$ yields unambiguous results.

In non-interacting case, $v_c=0$ is a physical choice because it satisfies the Euler equation for cold DM peculiar velocity.
\begin{equation}
v_c'+\mathcal{H}v_c=0\nonumber
\end{equation}
However, in the coupled case this point should be carefully investigated. The most generic equation of motion for cold DM reads\cite{31},
\begin{equation}
v_c'+\mathcal{H}v_c=-\frac{a^2{Q_{c}}^0}{\rho_c}v_c+\frac{a^2\delta Q_{pc}}{\rho_c}\label{cdmvelocity}.
\end{equation}
Compared with non-interacting case, two additional terms appear on the RHS  of above equation. The first term refers to the inertial force
density produced by the varying rest mass of cold DM in the system and the second term refers to the non-gravitational force density. The
non-gravitational force consists of two parts,
\begin{eqnarray}
\delta Q_{pc}=\delta Q_{pc}^I\mid_{t}+Q_{c}^0v_{t}\quad .
\end{eqnarray}
One is the viscosity $Q_{c}^0v_{t}$ due to the momentum transfer in  DM and DE, where $v_{t}$ is the average energy transfer velocity; another
one is external non-gravitational force $\delta Q_{p\lambda}^I\mid_{t}$ due to the collision effect at the early time of the universe. If we
neglect the non-gravitational force $\delta Q_{pc}$ in Eq.~(\ref{cdmvelocity}),  cold DM particles only suffer the attraction of gravity without
other external non-gravitational force. Eq.~(\ref{cdmvelocity}) still has ``free falling" solution $v_c=0$ which is the same as the
non-interacting case. Setting $v_c=0$, with a completely specified gauge condition, the synchronous gauge is as valid as any other gauge.

The choice of Synchronous gauge has a very close tie with the weak equivalence principle.  The Synchronous coordinate should be chosen to rest
upon the local inertial frame where the four acceleration $A=\psi=0$ of observer is zero.  As discussed above, if there is only gravitational
forces acting on cold DM bulk, the cold DM particles are still ``free falling" and the Synchronous gauge is valid for the cold DM frame which, in
turn, means that cold DM frame is a local inertial frame. Since the weak equivalence principle is valid in the local inertial frame, it should be
valid in the cold DM frame.

\end{document}